# Abrupt Rabi oscillations in a superoscillating electric field

D. G. Baranov,[1,2] A. P. Vinogradov,[1,2,3] and A. A. Lisyansky[4,*]

[1]*Moscow Institute of Physics and Technology, 9 Institutskiy per., Dolgoprudny 141700, Russia*
[2]*All-Russia Research Institute of Automatics, 22 Sushchevskaya, Moscow 127055, Russia*
[3]*Institute for Theoretical and Applied Electromagnetics, 13 Izhorskaya, Moscow 125412, Russia*
[4]*Department of Physics, Queens College of the City University of New York, Queens, NY 11367, USA*
*Corresponding author: lisyansky@qc.edu

We study counterintuitive dynamics of a two-level system (TLS) interacting with electric field superoscillating in time. We show that a TLS may be excited by an external light pulse whose spectral components are below the absorption line of the TLS. We attribute this unique dynamics to the Rabi oscillations of the TLS in a superoscillating driving field.

As was shown by Berry [1] in 1994, a band-limited function $f(t)$, i.e. the function whose Fourier transform satisfies $\hat{f}(\omega) = 0$ for all frequencies $|\omega| > \omega_{\max}$, may oscillate with a frequency greater than $\omega_{\max}$. Such functions were called superoscillatory (SO) functions. Since then, mathematical properties of SO functions have been studied in detail [2-4] and various approaches to their construction have been suggested [5, 6]. Superoscillations have also been discussed in the context of quantum mechanical wave functions [7-9]. In particular, it was predicted [7] that a particle with a low momentum obtains a large momentum after passing through a slit if the superoscillating region of the wave function overlaps with the slit. SO field patterns were employed to construct field distributions with subwavelength resolution [10-15] without the use of near field. A microscope with the subwavelength resolution operating in the far field has also been demonstrated [16]. Some similarities between the phenomenon of superoscillations and speckle patterns can be established [17]. In all these studies, spatial oscillations are the object of interest. At the same time, the mathematical descriptions of SO functions are identical in both spatial and time domains, so that one may consider responses of different physical systems to superoscillating external perturbations. An example of such interactions, which we study here, is the dynamics of a TLS placed in an external SO electromagnetic field.

A TLS in its ground state subjected to an oscillating electric field can be excited if the frequency of the oscillating field lies within the narrow absorption band of the TLS $\omega_0 - \Gamma < \omega < \omega_0 + \Gamma$ [18, 19]. Otherwise, the driving field interacts weakly with the TLS unless the high intensity of the field causes multiphoton electronic transition [20]. We address the intriguing question whether a TLS can be excited by a superoscillating electromagnetic field, whose spectral components are outside of the absorption band of the TLS.

In this paper, we study the dynamics of a TLS, which models a real emitter, driven by a strong superoscillating electric field. We demonstrate that, in principle, by proper tailoring of the driving field, the TLS can be inversely populated by an external optical pulse which all spectral components of the electric field fall below the TLS absorption band. We show that this phenomenon cannot be associated with multiphoton absorption.

The simplest way to construct a SO function is to directly synthesize a quickly varying (in terms of maxima and minima repetition rate) function from low frequency components (see, e.g., Ref. [21]). The SO function is sought in the form of a superposition of $N$ harmonic oscillations:

$$s(t) = \operatorname{Re} \sum_{n=1}^{N} c_n g_n(t), \quad g_n(t) = \exp(i\omega_n t), \tag{1}$$

where $|\omega_n| < \omega_{\max}$. To obtain SO behavior for this function, we choose a sequence of times $t_n$, $n = 1,...,N$, to which certain values $s(t_n) = s_n$ are assigned. Then coefficients $c_n$ are obtained by solving the system of linear equations:

$$s(t_j) = \sum_{n=1}^{N} c_n \exp(i\omega_n t_j) = s_j. \tag{2}$$

If the values $s_n$ are chosen such that they oscillate faster than the harmonic with the maximum frequency $\omega_{\max}$, then the function $s(t)$ will show SO behavior in the time interval $t_1 < t < t_N$.

In order to reveal the mechanism of the excitation process of a TLS in the SO field, we study the dynamics of the system during a short pulse. To do this, we multiply the SO function by the exponential factor which results in shaping of the pulse:

$$f(t) = s(t) \exp\left(-(t-t_0)^2/T^2\right), \tag{3}$$

where $t_0$ is the center of the pulse and $T$ denotes its characteristic duration. Of course, considering a slowly-varying pulse instead of a harmonic oscillation inevitably causes broadening of the signal spectrum. As a consequence, a certain part of the $f(t)$ spectrum overlaps with the absorption band of the TLS. However, as we show below, a non-resonant excitation of the TLS cannot be attributed to this broadening, provided that the pulse duration, $T$, is large enough.

The object of our interest is a quantum emitter driven by an external oscillating electric field. We model the emitter as a TLS with ground and excited states denoted by $|g\rangle$ and $|e\rangle$, respectively. The electric field is treated classically, i.e., it is assumed that the time dependence of the electric field amplitude is given by the predetermined function $f(t)$. This assumption is justified by the fact that the electric field is highly intensive, so that quantum effects on the electric field evolution may be neglected. The dynamics of a driven TLS is governed by the time-dependent Schrödinger equation with the usual Hamiltonian

$$\hat{H}_0 = \hbar\omega_0 \hat{\sigma}^\dagger \hat{\sigma} - \Omega f(t)(\hat{\sigma}^\dagger + \hat{\sigma}), \tag{4}$$

where $\omega_0$ is the TLS transition frequency, $\hat{\sigma} = |g\rangle\langle e|$ and $\hat{\sigma}^\dagger = |e\rangle\langle g|$, $\Omega = \mu E$ is the Rabi constant describing coupling of the TLS and the electric field, and $f(t)$ is the real-valued electric field shape function. Below we measure all frequencies in the units of $\omega_0$ and all times in the units of the inverse transition frequency $\omega_0^{-1}$. To take spontaneous decay into account, and dephasing of the TLS, we use the

density matrix operator $\hat{\rho}$. In the framework of the density matrix operator formalism, the equations governing the system evolution are [22]:

$$\dot{\rho}_1 = \rho_2 - \rho_1/T_2,$$
$$\dot{\rho}_2 = -\rho_1 - \rho_2/T_2 + 2\Omega f(t)\rho_3, \quad (5)$$
$$\dot{\rho}_3 = -2\Omega f(t)\rho_2 - (\rho_3 - \rho_{30})/T_1.$$

In Eqs. (5), terms $\rho_{1,2,3}$ are expressed through the elements of the density matrix as $\rho_1 = \rho_{12} + \rho_{21}$, and $\rho_3 = \rho_{22} - \rho_{11}$. In order to take into account the effects of dissipation, the phenomenological diagonal terms corresponding to spontaneous decay and dephasing processes with characteristic times $T_1$ and $T_2$, respectively, are included into Eqs. (5). The term $\rho_{30}$ represents the population inversion of the TLS at rest and it can be assumed that $\rho_{30} = -1$ at room temperature, i.e., the TLS decays to its ground state when not acted upon by an electric field. We assume that the dimensionless decay and dephasing times are equal to $T_1 = 500$ and $T_2 = 300$, respectively, and the Rabi constant to $\Omega = 0.01$.

For components of the band-limited SO function $s(t)$ we choose a basis of five harmonic oscillations $g_n(t) = e^{i\omega_n t}$ with frequencies $\omega_n = 0.18n$, $n = 1,...,5$. All harmonics from this set satisfy the condition $|\omega_n| < \omega_0$. With this basis of harmonic oscillations, we intend to obtain a SO function that oscillates locally during a certain time interval with the frequency of the TLS transition $\omega_0$, which is larger than the maximum frequency presented in the spectrum. To do this, we choose five points $t_n = \pi n/\omega_0$, $n = 0,...,4$, to which the following values are assigned: $s_1 = s_3 = s_5 = -1$, $s_2 = s_4 = 1$ [see the pattern depicted in Fig. 1(a)]. Solving the linear system of Eqs. (2), we obtain the desired SO function $s(t)$ which is plotted in Fig. 1(b). For the given pattern $s_n$, one obtains complex amplitudes for the harmonics as follows:

$$c_1 \approx -0.156 + 0.331i, \quad c_2 \approx -0.862 - 1.042i,$$
$$c_3 \approx 2.341 - 0.601i, \quad c_4 \approx -0.502 + 2.634i, \quad (6)$$
$$c_5 \approx -1.820 - 1.322i.$$

Fig. 1(b) shows the SO function $s(t)$ and the harmonic having the highest frequency from the set $\{\omega_n\}$. One can see from this figure that $s(t)$ oscillates within the interval $0 < t < 4\pi$ with local frequency $\omega_{loc} = 1$.

Now, we investigate the dynamics of the TLS in the external pulse of the form $f(t) = As(t)\exp\left[-(t-t_0)^2/T^2\right]$ with the pulse duration $T = 100$ and position of the pulse center $t_0 = 200$. Here $A$ is a real-valued constant amplitude whose value is chosen in such a way that strong coupling between the TLS and the electric field is attained. The pulse shape and its spectral density $\hat{f}(v) = \left| \int f(t) e^{ivt} dt \right|$ are plotted in Figs. 1 (c) and (d). Five peaks in Fig. 1(d) represent five harmonic components $g_n(t)$ which are broadened due to the exponential factor $\exp\left[-(t-t_0)^2/T^2\right]$.

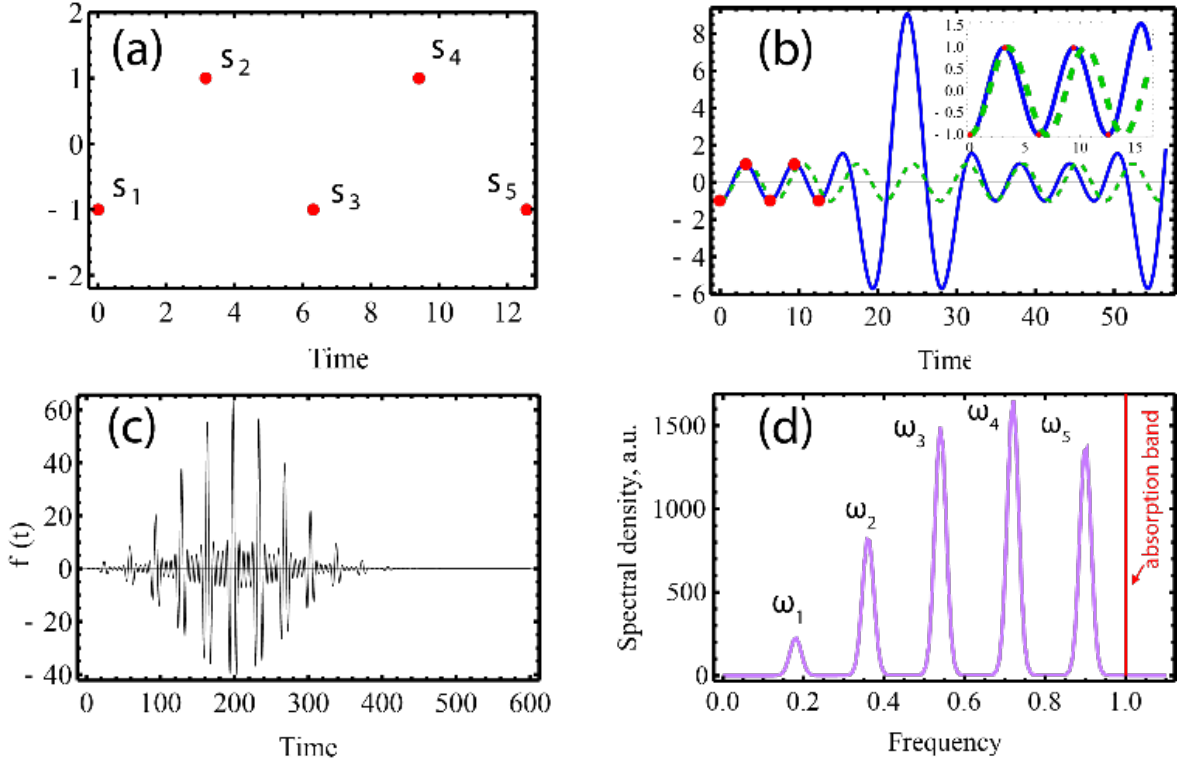

Fig. 1. (a) Predefined pattern of five points $\{t_n, s_n\}$ responsible for SO behavior of the function $s(t)$. (b) Solid line: the SO function $s(t)$ obtained by solving Eqs. (2) for the pattern depicted in panel (a). Dashed line: the fastest harmonic component of $s(t)$ having frequency $\omega_5 = 0.9$. The inset shows the time interval at which SO behavior is clearly visible. (c) The pulse shape function $f(t)$ with the duration $T = 100$ and the amplitude $A = 7$. (d) Normalized spectral density of pulse depicted in panel (c). Five peaks correspond to five spectral components of the initial SO function $s(t)$. Thin red line shows the absorption band of the TLS defined as $\omega_0 - \Gamma < \omega < \omega_0 + \Gamma$, with $\omega_0 = 1$ and $\Gamma = 1/500$. Peaks of electric field pulse do not overlap with the absorption band.

We first present the dynamics of the TLS driven by the field $f(t)$. The time dependence of the TLS-field coupling $\Omega f(t)$ and the population inversion $\rho_3$ are shown in Fig. 2 for three different values of the pulse amplitude $A = 1$, $A = 4$ and $A = 7$ (we chose these values of external pulse amplitudes as they clearly illustrate the dynamics of the TLS).

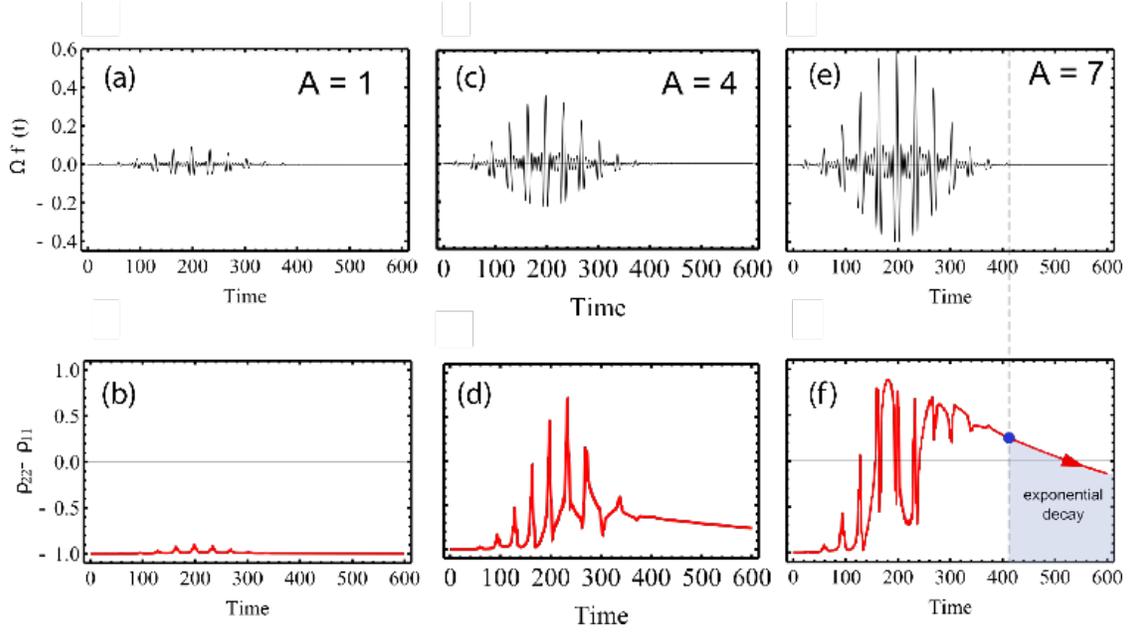

Fig. 2. (a, b, c) Time-dependent Rabi frequency $\Omega f(t)$ for a superoscillating pulse of three different amplitudes $A = 1, 4,$ and $7$. (d, e, f) Population inversion of the TLS driven by the field $f(t)$. For $A = 7$, the inverted population $\rho_{22} - \rho_{11} > 0$ is attained at the end of the superoscillating pulse.

Figure 2 represents the result central to our paper. When the TLS-field coupling $\Omega f$ is small (left panels, $A = 1$), the TLS almost does not interact with the external field [Fig. 2(d)]. At the middle panels, which correspond to the amplitude $A = 4$, one can observe population oscillations. Within the duration of the pulse, the population inversion may become positive, as indicated in Fig. 2(e). However, when the pulse action is finished, the population inversion returns nearly to its initial value $\rho_{30} = -1$, despite the superoscillating behavior of the pulse.

The right panel shows the most interesting result. In fact, it shows that a TLS can be excited by a non-resonant external electromagnetic field. At the end of the pulse, the TLS is excited with the inverted population $\rho_{22} - \rho_{11} > 0$. When the pulse action is finished ($t > 400$), the TLS experiences the usual exponential decay to its ground state. We claim that this unique non-resonant process of excitation is the direct manifestation of superoscillations of the external field.

Let us illustrate the significance of superoscillations for observation of non-resonant excitation. Consider evolution of the TLS population inversion in the electric field pulse

$f_5(t) = A \cos(\omega_5 t) \exp\left[-(t-t_0)^2 / T^2\right]$ formed by a *single harmonic* having the highest frequency $\omega_5$ from the set $\{\omega_n\}$. The amplitude $A = 25$ of such a pulse is chosen in such a way that the characteristic Rabi frequency $\Omega f_5$ during the pulse is of the same order as for the SO pulse for which the inverted population is attainable. Results presented in Fig. 3 clearly indicate that use of a single non-resonant harmonic does not allow one to achieve a positive population inversion. At the end of the pulse (time position $t = 400$), the value of the population inversion is $\rho_3 \approx -0.7$. For even stronger pulses, the resulting TLS population inversion is of the same order. No positive population inversion was observed in numerical simulation with a pulse of frequency $\omega_5$.

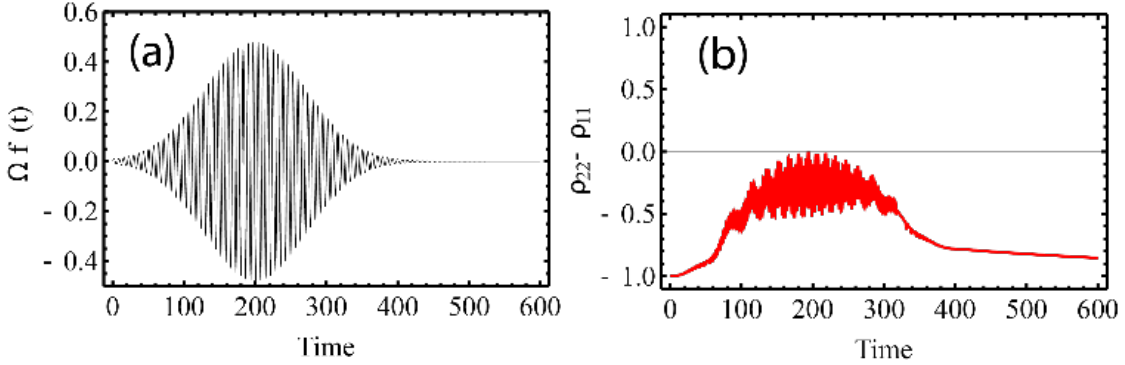

Fig. 3. Illustration of a non-resonant light-TLS interaction without superoscillations.
(a) The Rabi frequency $\Omega f_5(t)$ of a pulse formed by a single harmonic with frequency $\omega_5$. Characteristic value of the Rabi frequency is comparable to that observed in Fig. 1(e).
(b) Corresponding dynamics of the TLS population inversion $\rho_3 = \rho_{22} - \rho_{11}$.

The observed behavior of a TLS in a superoscillating pulse cannot be attributed to spectrum broadening. To justify this, we consider evolution of the TLS in the pulse of the same duration $T = 100$ formed by the *resonant* harmonic of frequency $\omega_0$:

$$f_{res}(t) = A_{res} \operatorname{Re} \exp(i\omega_0 t) \exp\left(-(t-t_0)^2 / T^2\right). \tag{7}$$

In this case, we set the value of amplitude $A_{res} = 0.02$. The spectral densities of such a resonant pulse and the SO pulse for which we observe inverted population, are shown in Fig. 4(a). Spectral densities of these two pulses are of the same order near the absorption region of the TLS. In Fig. 4(b), we plot the corresponding dynamics of the TLS population inversion coupled to the pulse of electric field (7). The resulting population inversion during the pulse is negligible in comparison with the values shown in Figs. 2 for the case of the SO pulse. In fact, the maximum value of the population inversion attainable during the action of a weak resonant pulse exceeds the "ground" (non-excited) inversion by of the order of $10^{-3}$. This dependence clearly indicates that the interaction of the TLS transition with the resonant part of the spectrum due to broadening is completely insufficient for achieving pronounced population inversion. Therefore, the observed dynamics of the TLS can only be associated with the superoscillating behavior of an electric field.

The predicted effect represents Rabi oscillations of a TLS in an external field. In our case, unlike usual Rabi oscillations, frequencies of external field harmonics are smaller than the TLS excitation frequency. However, thanks to the formation of superoscillations on a certain time interval, the external field oscillates with the characteristic period corresponding to the period of the TLS oscillations. The pulse envelope is chosen so that Rabi oscillations are stopped when the TLS is excited. If the duration of the pulse is slightly longer, the TLS may return to the ground state, similar to the usual Rabi oscillations. Further increase of the pulse again leads to the excitation of the TLS. This process is repeated with an increase of the pulse duration.

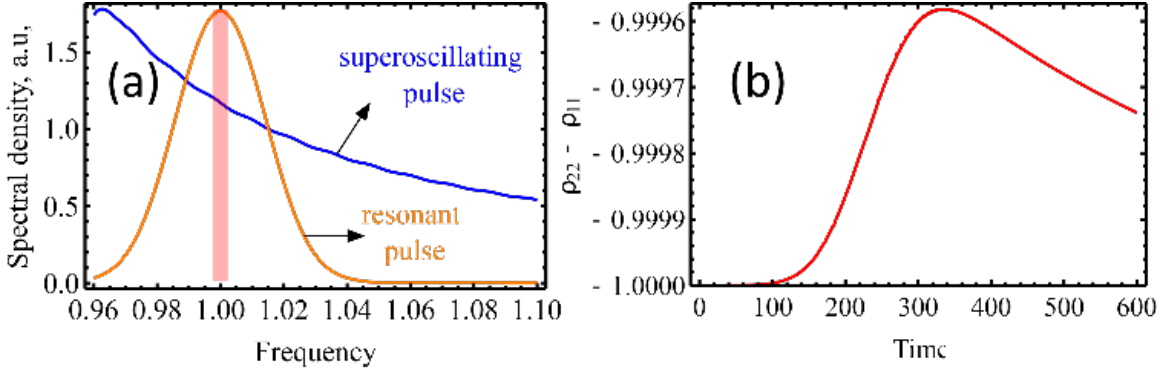

Fig. 4. Dynamics of a TLS in a weak resonant electric field pulse (7). (a) Spectral densities of the SO pulse and the resonant pulse. The red shaded region indicates the absorption band of the TLS. (b) Corresponding dynamics of the TLS interacting with the resonant pulse $f_{res}(t)$.

It is worth emphasizing that there are some fundamental limitations on superoscillations which could make their use problematic in practice [3, 23]. The main obstacle in the construction of a SO function is that the $L_2$- norm of the resulting function increases polynomially with the desired frequency of superoscillations and exponentially with the number of oscillations. As a result, the amplitude of high frequency oscillations becomes small in comparison with that of the low-frequency region. In other words, the energy of a SO signal is concentrated in the lower frequency region.

It may seem that a high intensity field may cause the multiphoton excitation of the TLS. We show that this is not so. The phenomenon demonstrated here cannot be explained by a multiphoton absorption. Frequencies $\omega_n$ making up the basis of harmonic oscillations $g_n(t)$ in Eq. (1) are arbitrary. The only condition which the basis set $g_n(t)$ must satisfy is the solvability of Eqs. (2). Thus, we can choose such a set of frequencies that for any pair of frequencies $\omega_k$ and $\omega_l$ the condition of the two-photon absorption $\omega_k + \omega_l = \omega_0$ does not hold. This is the case for the particular set of frequencies $\{\omega_n\}$ that we use in present calculations.

The TLS studied here is a great simplification of the energy structure of a real atom. Atomic structures may contain transitions whose frequencies are close to components $\omega_n$ of the SO electric field. Therefore, those transitions will strongly interact and absorb lower-frequency photons, which play crucial role in superoscillating behavior of the electric field. Nevertheless, as noted in the previous paragraph, the basis frequencies are to some extent arbitrary. Thus, it allows one to tailor the set of frequencies $\omega_n$ in such a way that none of these frequencies is attributed to absorption bands of a certain energy structure of an atom.

It is important to note that the Stark effect is taken into account in Hamiltonian (4). Therefore, the predicted excitation of the TLS occurs is not hampered by the Stark shift in the energy levels. Indeed, in the second order of the perturbation theory, the Stark effect shifts both the ground $E_g$ and excited $E_e$ state energies of the TLS by $\Delta E_g = |\Omega f|^2 / (E_g - E_e)$ and $\Delta E_e = |\Omega f|^2 / (E_e - E_g)$, respectively [24]. Since $E_e > E_g$, the resulting frequency shift is always *positive*: $\delta \omega = (\delta E_e - \delta E_g)/\hbar \approx 0.5\omega_0$. At the same time, spectral components of the external electric pulse have frequencies *below* the transition frequency of the TLS.

In conclusion, we have investigated the unique temporal dynamics of a quantum emitter driven by an external superoscillating field. It is possible to construct a superoscillating pulse of electromagnetic field, using only non-resonant harmonics, which can excite the emitter. By tailoring the shape and amplitude of the pulse we achieve population inversion of the emitter at the end of the superoscillating pulse. The observed dynamics is associated with the superoscillations phenomenon and cannot be attributed to multiphoton processes or spectrum broadening. This phenomenon shows the fundamental possibility of exciting atomic transitions by an electromagnetic pulse whose spectrum fall below the atomic transition line.

The work was supported by RFBR grants No 13-02-00407 and 13-07-92660, by Dynasty Foundation, and by the NSF under Grant No. DMR-1312707.

References
1. M. V. Berry, J. Phys. A-Math. Gen. **27**, 391 (1994).
2. A. Kempf, J. Math. Phys. **41**, 2360 (2000).
3. P. J. S. G. Ferreira and A. Kempf, IEEE T Sign Proces **54**, 3732 (2006).
4. Y. Aharonov, F. Colombo, I. Sabadini, D. C. Struppa, and J. Tollaksen, J. Phys. A-Math. Theor. **44**, 365304 (2011).
5. Q. Wang, J. Phys. A-Math. Gen. **29**, 2257 (1996).
6. P. J. S. G. Ferreira, A. Kempf, and M. J. C. S. Reis, J. Phys. A-Math. Theor. **40**, 5141 (2007).
7. A. Kempf and P. J. S. G. Ferreira, J. Phys. A-Math. Gen. **37**, 12067 (2004).
8. M. S. Calder and A. Kempf, J. Math. Phys. **46**, 012101 (2005).
9. M. V. Berry and S. Popescu, J. Phys. A-Math. Gen. **39**, 6965 (2006).
10. F. M. Huang, N. Zheludev, Y. Chen, and F. Javier Garcia de Abajo, Appl. Phys. Lett. **90**, 091119 (2007).
11. F. M. Huang, Y. Chen, F. J. Garcia de Abajo, and N. I. Zheludev, J. Opt. A **9**, S285 (2007).
12. F. M. Huang, T. S. Kao, V. A. Fedotov, Y. Chen, and N. I. Zheludev, Nano Lett. **8**, 2469 (2008).
13. F. M. Huang and N. I. Zheludev, Nano Lett. **9**, 1249 (2009).
14. K. G. Makris and D. Psaltis, Opt. Lett. **36**, 4335 (2011).

15. E. T. F. Rogers, S. Savo, J. Lindberg, T. Roy, M. R. Dennis, and N. I. Zheludev, Appl. Phys. Lett. **102**, 031108 (2013).
16. E. T. F. Rogers, J. Lindberg, T. Roy, S. Savo, J. E. Chad, M. R. Dennis, and N. I. Zheludev, Nat. Mater. **11**, 432 (2012).
17. M. R. Dennis, A. C. Hamilton, and J. Courtial, Opt. Lett. **33**, 2976 (2008).
18. P. Meystre, Atom Optics (Springer-Verlag New York, 2001).
19. L. Allen and J. Eberly, Optical Resonance and Two-Level Atoms (Courier Dover Publications, 1978).
20. R. Loudon, The Quantum Theory of Light (Clarendon Press, 1973).
21. K. G. Makris and D. Psaltis, "Superoscillatory diffraction-free beams," Opt. Lett. 36, 4335 (2011).
22. R. W. Ziolkowski, J. M. Arnold, and D. M. Gogny, Phys. Rev. A 52, 3082 (1995).
23. H. J. Hyvärinen, S. Rehman, J. Tervo, J. Turunen, and C. J. R. Sheppard, Opt. Lett. **37**, 903 (2012).
24. L. D. Landau and E. M. Lifshitz, Quantum Mechanics (Butterworth-Heinemann, 1976).